\documentstyle[preprint,epsbox]{jpsj}

\newcommand{\LSIM}{\mbox{\raisebox{-0.5ex}{$\stackrel{<}{\sim}$}}}

\def\runtitle{Collisionless Damping of Low-Frequency Magnetosonic Pulses in a
Two-Ion-Species Plasma}
\def\runauthor{Daiju {\sc Dogen}, Mieko {\sc Toida} and Yukiharu {\sc Ohsawa}}

\title
{
Collisionless Damping of Low-Frequency Magnetosonic Pulses in a
Two-Ion-Species Plasma
}

\author
{ 
Mieko {\sc Toida}, Daiju {\sc Dogen} and Yukiharu {\sc Ohsawa}
}

\inst
{
 Department of Physics,
Nagoya University, Nagoya 464-8602.
}

\recdate
{
\today
}

\abst
{Low-frequency mangnetosonic pulses in a two-ion-species plasma are studied theoretically and by simulation with a one-dimensional electromagnetic simulation code based on a three-fluid model, with particular attention to 
the dynamics of minority heavy ions.
It is found that heavy ions can gain some energy from the pulses.  
Because of this energy transfer, the  pulses are damped even if the
plasma is collisionless and pulse propagation is perpendicular to the magnetic field.}
\kword
{ collisionless damping, magnetosonic wave, two-ion-species plasma, 
energy dissipation, soliton
}

\begin{document}
\sloppy
\maketitle

The presence of multiple ion species introduces many interesting effects 
to linear and nonlinear magnetosonic waves.$^{1-9)}$
First, the magnetosonic wave in a two-ion-species plasma
 is split into two modes:
the high- and low-frequency modes.\cite{rf:twoKdV}
The frequency of the low-frequency mode tends to zero as the wave number $k$
tends to zero and approaches the ion-ion hybrid resonance frequency as $k \to
\infty$.\cite{rf:Buch} 
The high-frequency mode has a cut-off frequency on the order of ion
cyclotron frequency and has a resonance frequency on the order of the lower
hybrid frequency.
(The dispersion curves can be found in ref .1).
It was found that the nonlinear behavior of these modes are both described 
by the Korteweg-de Vries (KdV) equation, even though their dispersion curves are quite different in the long-wavelength region.\cite{rf:twoKdV}
The characteristic soliton widths are the electron skin depth for the high-frequency mode and the ion skin depth for the low-frequency mode.
\par
\indent
In a collisionless, single-ion-species plasma, small-amplitude magnetosonic waves propagating perpendicular to a magnetic field are not damped. \cite{rf:Bern, rf:Gard} (Large-amplitude waves can be damped, because they accelerate a fraction of the ions by the electrostatic field.
$^{12-15)}$)  It was reported in refs. 4 and 5, however, that a nonlinear pulse of the high-frequency mode is damped in a multi-ion-species plasma.
(Here, light ions are assumed to be the main component, as in space plasmas.)
The damping is due to the energy transfer to heavy ions. 
That is, the transverse electric field in a high-frequency-mode pulse can 
 accelerate heavy ions in the direction parallel to the wavefront.\cite{rf:heavac}. Thus the kinetic energy of heavy ions is increased behind the pulse.
 This could be an important dissipation mechanism in a
collisionless multi-ion-species plasma such as the solar corona.
\par
\indent
In this letter, we will show that the low-frequency mode pulse can also impart some energy to heavy ions in the same way.
Hence the pulse can be damped. An important difference is, however, that the heavy ions can suffer cyclotron oscillation a few times while they are in the pulse region,
because the soliton width of the low-frequency mode is quite large, about 
$(m_{\rm i}/m_{\rm e})^{1/2}$ times as large as that of the high-frequency mode, where $m_{\rm i}$ is the ion mass and $m_{\rm e}$ is the electron mass.
The heavy ions therefore alternately gain and lose energies in the pulse region. Here, we will obtain the net change in the kinetic energy of the heavy ions and study the wave damping due to this mechanism. 
First, we will analytically discuss the heavy-ion motion in the low-frequency 
mode pulse. 
It is found that heavy ions can be slightly accelerated by the pulse.
Next, using a simulation based on a three-fluid model,
we will show that the low-frequency-mode pulse imparts some energy to 
heavy ions as theoretically predicted. Therefore, the pulse is gradually damped.
\par
\indent
We consider magnetosonic waves propagating perpendicular to a magnetic field
in a plasma containing two ion species; they are denoted by {\it a} and $b$, and the ion cyclotron frequency of ions $b$ is assumed to be lower than that of $a$ 
$(\Omega_b < \Omega_a)$.
In a two-ion-species plasma, the dispersion relation of the low-frequency mode 
is given by
\begin{equation}
\omega = v_{\rm p0} k (1 - k^2 d^2/2 )
\label{eqn:lowdisp}
\end{equation}
in the long-wavelength region.
Here  $v_{\rm p0}$ is defined as
\begin{equation}
v_{\rm p0} = v_{\rm A} /(1 + v_{\rm A}^2/c^2)^{1/2},
\label{eqn:vpodef}
\end{equation}
where $v_{\rm A}$ is the Alfv\'en speed, $v_{\rm A}=B_0/(4 \pi \rho_0)^{1/2}$ with $\rho_0$ the average mass density 
$\rho_0 = n_{a0} m_a + n_{b0} m_b$.
The length $d$ in eq. (\ref{eqn:lowdisp}) is defined as
\begin{equation}
d = \frac{v_{\rm p0}^3}{c^2} \left[
\frac{\omega_{{\rm p}a}^2 \omega_{{\rm p}b}^2}{\Omega_a^2 \Omega_b^2}
\left( \frac{1}{\Omega_a} - \frac{1}{\Omega_b} \right)^2
+
\frac{\omega_{{\rm p}b}^2 \omega_{\rm pe}^2 }{\Omega_b^2 \Omega_{\rm e}^2}
\left( \frac{1}{\Omega_b} - \frac{1}{\Omega_{\rm e}} \right)^2
+
\frac{\omega_{\rm pe}^2 \omega_{{\rm p}a}^2}{\Omega_{\rm e}^2 \Omega_a^2}
\left( \frac{1}{\Omega_{\rm e}} - \frac{1}{\Omega_a} \right)^2
+ \sum_j \frac{\omega_{{\rm p}j}^2}{\Omega_j^4}
\right]^{1/2},
\label{eqn:dminus}
\end{equation}
where $\Omega_{\rm e}$ is negative and $\omega_{{\rm p}j}$ is the plasma 
frequency of particle species {\it j}.
\par
\indent 
When two ion species are present at considerable densities, the first term in
the square brackets, which is proportional to $(\Omega_a^{-1} -
\Omega_b^{-1})^2$, is the dominant term, and $d$ is on the order of the ion
skin depth,
$c/\omega_{{\rm pi}}$. The dispersion of the low-frequency mode is $\sim
(m_{\rm i}/m_{\rm e})$
times as large as that of the magnetosonic wave in a single-ion-species 
plasma.\cite{rf:Mikh}
In the limit of $n_{b0} \to 0$, the first and second terms disappear, and the electron inertia effect appearing in the third term becomes important.
The displacement currents produce the fourth term $\sum_j \omega_{{\rm p}j}^2/\Omega_j^4$, which is about $\Omega_{\rm pe}^2/\omega_{\rm pe}^2$ times as 
large as the third term.

\par
\indent
As can be expected from eq. (\ref{eqn:lowdisp}), the nonlinear low-frequency
wave can be described by the KdV equation,
\begin{equation}
\frac{\partial B_1}{\partial \tau}
+ \frac{3}{2} \frac{ v_{\rm p0}^2}{v_{\rm A}} \frac{B_1}{B_0}
\frac{\partial B_1}{\partial \xi} + \frac{1}{2} v_{\rm p0} d^2
\frac{\partial^3 B_1}{\partial \xi^3} =0,
\label{eqn:lowkdv}
\end{equation}
where $\xi$ and $\tau$ are stretched coordinates
\begin{equation}
\xi = \epsilon^{1/2} (x - v_{\rm p0} t),
\label{eq:xi}
\end{equation}
\begin{equation}
\tau = \epsilon^{3/2} t,
\label{eq:tau}
\end{equation}
with $\epsilon$ being the smallness parameter on the order of the amplitude
of, for instance, the magnetic field:
\begin{equation}
B_z = B_0 + \epsilon B_1 + \epsilon^2 B_2 + \cdots.
\label{eq:expand}
\end{equation}
Here, it was assumed that the waves propagate in the {\it x} direction and the magnetic field is in the {\it z} direction.
The magnetic field 
profile of the solitary wave can be given by
\begin{equation}
B_{z}(x,t) = B_0 \left[ 1 + B_{\rm n}{\rm sech}^{2}\left( \frac{x-Mv_{\rm
p0}t} {D} \right )  \right] .
\label{eqn:solitonBz}
\end{equation}
Here, $B_{\rm n}$ is the normalized wave amplitude, $B_{\rm n} = \epsilon B_1/B_0$, {\it D} is the soliton width,
\begin{equation}
D=2 (v_{\rm A}/v_{\rm p0}) d B_{\rm n}^{-1/2},
\label{eqn:solitonD}
\end{equation}
and {\it M} is the Mach number related to the wave amplitude $B_{\rm n}$
through
\begin{equation}
M = 1 + (v_{\rm p0}/v_{\rm A})^2 B_{\rm n}/2.
\label{eqn:solitonM}
\end{equation}
The longitudinal electric field $E_x$ and the transverse electric 
field $E_y$ have profiles
\\
\begin{equation}
E_{x}(x,t)  =  \frac{v_{\rm p0}^{5} B_0}{c^{3}v_{\rm A} d}\left(
\sum_{j}\frac{\omega_{{\rm p}j}^{2}}{\Omega_{j}^{3}}\right) B_{\rm n}^{3/2}{\rm
sech}^{2}\left(
\frac{x-Mv_{\rm p0}t}{D}\right){\rm tanh}\left( \frac{x-Mv_{\rm
p0}t}{D}\right),
\label{eqn:solitonEx}
\end{equation}
\begin{equation}
E_{y}(x,t)  =  \frac{v_{\rm p0}B_{0}}{c} B_{\rm n}{\rm sech}^{2}\left(
\frac{x-Mv_{\rm p0}t}{D}\right). 
\label{eqn:solitonEy}
\end{equation}
If $n_{b}$ is taken to be zero, then the above KdV theory is reduced to the one
for a single-ion-species plasma.
\par
\indent
We now discuss single-particle motion of heavy ions in the solitary pulse using the equations of motion: 
\begin{equation}
m_b \frac{\mbox{d} v_{bx}}{\mbox{d} t} = q_b \left[ E_x(x,t) + \frac{v_{by}}{c} B_z(x,t) \right],
\label{eqn:vbxMotion}
\end{equation}
\begin{equation}
m_b \frac{\mbox{d} v_{by}}{\mbox{d} t} = q_b \left[ E_y(x,t) - \frac{v_{bx}}{c} B_z(x,t) \right].
\label{eqn:vbyMotion}
\end{equation}
We note that the relation $E_y - v_x B_z/c =0$ holds for both ions and electrons in nonlinear magnetosonic waves in a single-ion-species plasma.\cite{rf:Adlam,rf:Davis} 
Thus $E_y$ does not accelerate ions in such a case.
The KdV theory for the low-frequency mode also gives the relation 
$v_{bx} =  c E_{y}/B_0$ among the lowest-order perturbations.\cite{rf:twoKdV} 
The discussion in ref. 3, however, shows that this relation can easily break down for heavy ions. 
Here we calculate, more accurately, the heavy-ion motion, including inertial effects, and examine if the heavy ions can gain energies.
\par
\indent
We integrate eqs. (\ref{eqn:vbxMotion}) and (\ref{eqn:vbyMotion}) over time, 
assuming that the initial particle velocity is zero, 
$v_{bx}(0) = v_{by}(0) = 0$, and the initial particle position is in the
far upstream region, $x(0) \equiv x_0 \gg D$.
We disregard the perturbation of the
magnetic field and approximate the variable $x$ in $E_x$ and $E_y$ as $x=x_0$.
We take the time derivative of eq. (\ref{eqn:vbyMotion}) and eliminate $v_{bx}$ using eq. (\ref{eqn:vbxMotion}), and obtain
\begin{equation}
\frac{{\rm d}^{2} v_{by}}{{\rm d}t^{2}}  + \Omega_{b}^{2} v_{by} =  F(t).
\label{eqn:vbyDifeq}
\end{equation}
Here $F(t)$  is defined as
\begin{equation}
F(t) = \frac{\Omega_b v_{\rm p0}^3}{v_{\rm A} d}
\left[1 - \frac{v_{\rm p0}^2}{c^2} \left( \sum_j 
\frac{\omega_{{\rm p}j}^2}{\Omega_j^3} \right)
 \Omega_b \right] B_{\rm n}^{3/2}
\mbox{sech}^2 \left( \frac{x_0 -Mv_{\rm p0}t}{D} \right)
\mbox{tanh} \left( \frac{x_0 -Mv_{\rm p0}t}{D} \right).
\label{eqn:Ftdisp}
\end{equation}
We apply the Laplace transform to eq. (\ref{eqn:vbyDifeq}). Then, with the aid of the inversion formula, 
the velocity $v_{by}$ for $t > 0$ can be found as
\begin{equation}
v_{by}(t) = \frac{1}{\Omega_b}\int_{0}^{t} \sin [\Omega_b(t-u)] F(u) {\rm d}u.
\label{eqn:integvby}
\end{equation}
We can also have $v_{bx}$ as 
\begin{equation}
v_{bx}(t) = - \frac{1}{\Omega_b}\int_{0}^{t} \cos [\Omega_b (t-u)] F(u) {\rm d}u 
+ v_{\rm p0} B_{\rm n} {\rm sech}^{2}\left(
\frac{x_0 -Mv_{\rm p0}t}{D}\right). 
\label{eqn:integvbx}
\end{equation}
To obtain the velocity $v_{by}$ in the downstream region 
at large $t$, we assume that
$(x_0 - Mv_{\rm p0} t)/D \to  -\infty$.
Then, $v_{by}$ can be obtained as
\begin{equation}
v_{by} =  v_{bm} \cos \left[ \Omega_b \left( t - \frac{x_{0}}{Mv_{\rm p0}}\right)
\right ], 
\label{eqn:vbySolution}
\end{equation}
where $v_{bm}$ is defined as
\begin{equation}
v_{bm} = 4 \pi \frac{\Omega_b^2 d^2 v_{\rm A}^2}{v_{\rm p0}^3}
{\rm cosech} \left( \frac{\pi \Omega_b d v_{\rm A}}{v_{\rm p0}^2
 B_{\rm n}^{1/2} [1 + (v_{\rm p0}^2/v_{\rm A}^2) B_{\rm n}/2] }
\right).
\label{eqn:vbmdisp}
\end{equation}

If we neglect the effects of the displacement currents and the 
electron inertia, $v_{bm}$ can be written as
\begin{equation}
v_{bm} = 4 \pi v_{\rm A} \frac{n_{b0}m_b n_{a0}^2m_a^2}{\rho_o^3}
 \left(1 - \frac{\Omega_b}{\Omega_a} \right)^3
{\rm cosech} \left( \frac{\pi (n_{a0}m_a n_{b0}m_b)^{1/2} (1 - \Omega_b/ \Omega_a) }{ \rho_0 B_{\rm n}^{1/2} (1 + B_{\rm n}/2)}
\right). 
\label{eqn:vbmSolution}
\end{equation}
Equation (\ref{eqn:vbySolution}) indicates that heavy ions gyrate with the speed $v_{bm}$  in the downstream region. 
Thus, heavy ions have finite speed $v_{bm}$ behind the pulse, even though their initial speed was assumed to be zero.
\begin{figure}[htp]
\centering
\epsfile{file=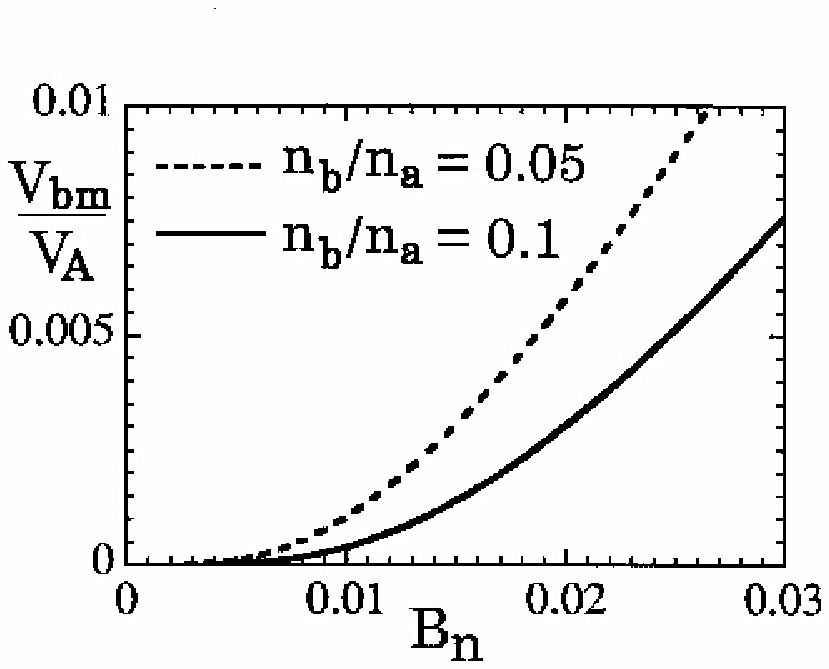}
\caption{Heavy-ion speed behind the pulse, $v_{bm}$, as a function of the wave 
amplitude $B_{\rm n}$.  The solid and dotted lines are for the plasmas 
with $n_{b}/n_{a} = 0.1$ and $n_{b}/n_{a} = 0.05$, respectively.}
\label{Fig:1}
\end{figure}
\par
\indent
Figure 1 shows $v_{bm}$ as a function of the wave amplitude $B_{\rm n}$ 
in a hydrogen-helium plasma where the mass and charge ratios are $m_b/m_a =4$ 
and $q_b/q_a =2$, respectively.
The solid line is for the plasma with the density ratio $n_{b}/n_{a}=0.1$ 
and the dotted line is for the plasma with $n_{b}/n_{a}=0.05$. 
The speed $v_{bm}$ increases with the wave amplitude $B_{\rm n}$ and decreases 
with the density ratio $n_{b}/n_{a}$.
\par
\indent
If we substitute the soliton solution for the high-frequency mode \cite{rf:twoKdV, rf:Jyou} in eqs. (\ref{eqn:vbxMotion}) and (\ref{eqn:vbyMotion}), the heavy-ion speed behind the high-frequency-mode pulse is obtained as
\begin{equation}
v_{bm} = 4 \pi v_{\rm h}  \alpha' \frac{\Omega_{b}^2}{\Omega_{\rm e}^2}
\frac{\omega_{\rm pe}^2 \omega_{{\rm p}a}^2}
{(\omega_{{\rm p}a}^2 + \omega_{{\rm p}b}^2 )^2 }
\left( 1 - \frac{\Omega_b}{\Omega_a} \right)  
\mbox{cosech} \left( \frac{\pi \Omega_b c \alpha'^{1/2}}{v_{\rm h} \omega_{\rm pe} B_{\rm n}^{1/2} ( 1 + B_{\rm n}/2\alpha') } \right) .
\label{eqn:vbmhigh}
\end{equation}
The speed $v_{\rm h}$ is slightly greater than $v_{\rm A}$ and $\alpha'$ is an order-unity quantity; for their precise expressions, see ref. 3.
In the limit that the transit time is much shorter than the ion cyclotron period, eq. (\ref{eqn:vbmhigh}) is reduced to the one derived in the previous paper,
 eq. (A.14) in ref. 3. 
\begin{figure}[htp]
\centering
\epsfile{file=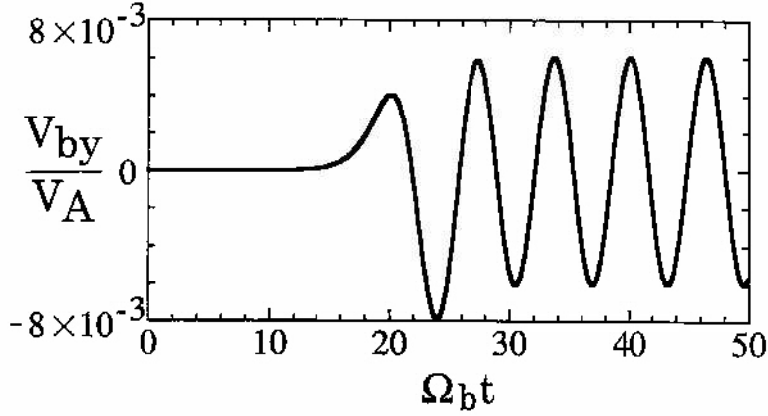}
\caption{Time variation of $v_{by}$  obtained by numerical 
integration of the equation of motion.}
\label{Fig:2}
\end{figure}
\par
\indent
Equation (\ref{eqn:vbySolution}) gives $v_{by}(t)$ in the far downstream region. To observe the time variation of $v_{by}$ in and outside the pulse region,
we numerically integrated eqs. (\ref{eqn:vbxMotion}) and (\ref{eqn:vbyMotion}), substituting the low-frequency-mode pulse given by eqs. (\ref{eqn:solitonBz})-(\ref{eqn:solitonEy}).
The amplitude of the pulse was taken to be $B_{\rm n}=0.03$.
Figure 2 shows the time variation of $v_{by}$ thus obtained.
When a heavy ion enters the pulse at time $\Omega_b t = 15$, 
$v_{by}$ starts to increase. Because the Lorentz force converts $v_{by}$ to $v_{bx}$, $v_{by}$ decreases after reaching the maximum value.
At about $\Omega_{b} t= 25$, the particle goes out of the pulse and 
starts to gyrate with the speed $v_{bm}= 6.0\times 10^{-3}v_{\rm A}$.

\par
\indent
Next, we discuss the energy change of a heavy-ion particle,
\begin{equation}
\Delta E_b = q_b \int_0^t E_x v_x {\rm d}t + q_b \int_0^t  E_y v_y {\rm d}t,
\label{eqn:DeltEb1}
\end{equation}
using eqs. (\ref{eqn:integvby}) and (\ref{eqn:integvbx}).
The energy gain from the low-frequency-mode pulse
can be written as

\begin{equation}
q_b \int_0^{\infty} E_x v_x {\rm d}t =  -\frac{m_b}{2} v_{bm}^2
\sum_j \frac{\omega_{{\rm p}j}^2}{\Omega_j^3} \Omega_b
 \left[ \frac{\omega_{{\rm p}a}^2}{\Omega_a^3}
(\Omega_a - \Omega_b) + 1 + \frac{\omega_{\rm pe}^2}{\Omega_{\rm e}^2} 
\right]^{-1},
\label{eqn:intExvx}
\end{equation}
\begin{equation}
q_b \int_0^{\infty} E_y v_y {\rm d}t =  \frac{m_b}{2} v_{bm}^2
\frac{c^2}{v_{\rm p0}^2} \left[ \frac{\omega_{{\rm p}a}^2}{\Omega_a^3}
(\Omega_a - \Omega_b) + 1 + \frac{\omega_{\rm pe}^2}{\Omega_{\rm e}^2} 
\right]^{-1}.
\label{eqn:intEyvy}
\end{equation}
The energy gain from the longitudinal electric field $E_x$ is
negative, and that from the transverse electric field $E_y$ is
positive. Hence, the heavy ions obtain energies from the 
transverse electric field. 
The net change in the energy is
\begin{equation}
\Delta E_b = m_b v_{bm}^2 /2.
\label{eqn:DeltEb2}
\end{equation}
It is thus expected that the pulse of the low-frequency mode is damped due
to this energy transfer.
\par
\indent
Let us now study the propagation of the low-frequency mode using a
one-dimensional, fully electromagnetic code based on the three-fluid model:
\begin{equation}
\frac{\partial n_{j}}{\partial t} +
\mbox{\boldmath{$\nabla$}}\cdot\left(n_{j}\mbox{\boldmath{$v$}}_{j}\right)=0,
\label{eqn:continuosEq}
\end{equation}
\begin{equation}
m_{j}\left[\frac{\partial}{\partial t} +
\left(\mbox{\boldmath{$v$}}_{j}\cdot\mbox{\boldmath{$\nabla$}}\right)\right]
\mbox{\boldmath{$v$}}_{j}=q_{j}\mbox{\boldmath{$E$}}+\frac{q_{j}}{c}
\mbox{\boldmath{$v$}}_{j}\times\mbox{\boldmath{$B$}},
\label{eqn:MotionEq}
\end{equation}
\begin{equation}
\frac{1}{c}\frac{\partial\mbox{\boldmath{$B$}}}{\partial t} =
-\mbox{\boldmath{$\nabla$}}\times\mbox{\boldmath{$E$}}, 
\label{eqn:FaradayEq}
\end{equation}
\begin{equation}
\frac{1}{c}\frac{\partial\mbox{\boldmath{$E$}}}{\partial t} =
\mbox{\boldmath{$\nabla$}}\times\mbox{\boldmath{$B$}}  -
\frac{4\pi}{c}\sum_{j}q_{j}n_{j}\mbox{\boldmath{$v$}}_{j} .
\label{eqn:AmperesEq}
\end{equation}
The $x$ component of eq. (\ref{eqn:AmperesEq}) gives the longitudinal electric field $E_x$.
The periodic boundary conditions are assumed. As the initial wave profiles, we use the solitary wave solutions obtained from the KdV equation for the
low-frequency mode, eq. (\ref{eqn:lowkdv}), and observe their evolution.
First, we confirmed that solitary pulses are not damped in a single-ion-species plasma ($n_b = 0$). 
We then simulated a hydrogen-helium plasma and chose the hydrogen-to-electron mass ratio as $m_a/m_{\rm e}=1000$. The density ratio was $n_b/n_a=0.1$, 
and the magnetic field strength was
$|\Omega_{\rm e}|/\omega_{\rm pe} = 0.5$, so that $c/v_{\rm A} =68.3$.
\begin{figure}[htp]
\centering
\epsfile{file=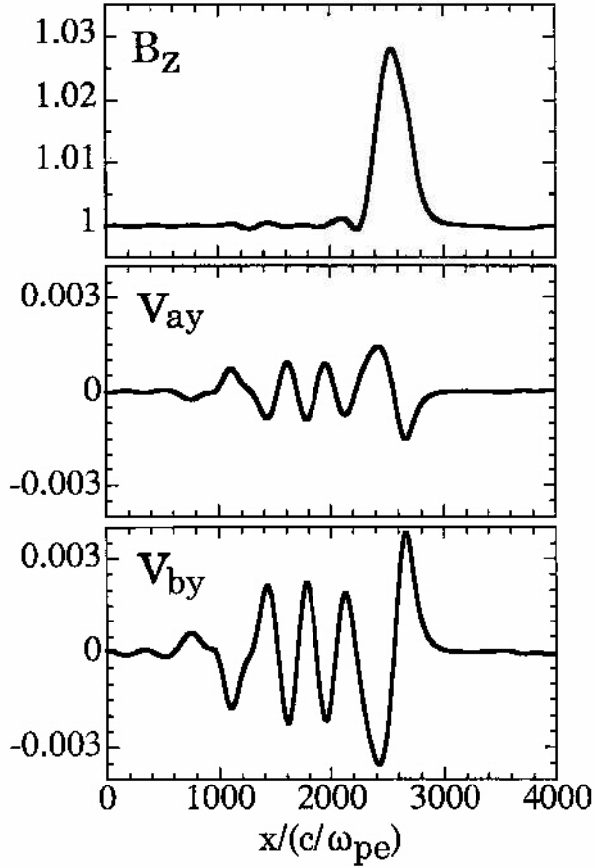}
\caption{Profiles of the magnetic field, light-ion velocity $v_{ay}$, and heavy-ion velocity $v_{by}$ at $\Omega_{b} t = 25$. The initial amplitude is $B_{\rm n} = 0.03.$}
\label{Fig:3}
\end{figure}
\par
\indent
In Fig. 3 we show profiles of the magnetic field, light-ion (H) velocity $v_{ay}$,
and heavy-ion (He) velocity $v_{by}$  at time $\Omega_b t = 25$.
The magnetic field and velocities are normalized by $B_0$ and $v_{\rm A}$, respectively.
 At time $t =0$, the amplitude was $B_{\rm n}(0) =0.03$, and the center of the pulse was at $x/(c/\omega_{\rm pe}) \simeq 1100$.
Figure 3 shows that the heavy ions have a finite speed 
behind the pulse, as predicted by eq. (\ref{eqn:vbySolution}) and Fig. 2.
This heavy-ion motion produces perturbations of other components 
behind the pulse, 
$ 1000  \LSIM  x/(c/\omega_{\rm pe})  \LSIM  2300$, which can be seen in the profiles of $v_{ay}$ and $B_z$; even in the profile of $B_z$, we notice small-amplitude perturbations behind the main pulse.

\par
\indent
The observed heavy-ion speed $v_{bm}$,
 $2.2 \times 10^{-3} v_{\rm A}$, 
is smaller than the theoretical value given by eq. (\ref{eqn:vbmdisp}),
$v_{bm}=7.7 \times 10^{-3} v_{\rm A}$.
In deriving eq. (\ref{eqn:vbmdisp}), we have not included the effect of the generation of hydromagnetic perturbations behind the pulse.
Therefore, eq. (\ref{eqn:vbmdisp}) is interpreted to give the upper limit 
of the velocity of accelerated heavy ions.
\par
\indent
Figure 4 shows the time variation of the wave energy $E_{\rm w}(t)$ 
of the main pulse. Even though we observe small-amplitude fluctuation,
 the main pulse certainly loses energy gradually.
\begin{figure}[htp]
\centering
\epsfile{file=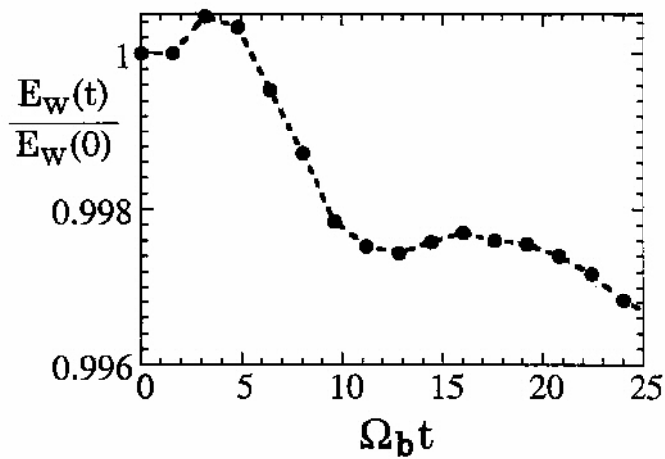}
\caption{Time variation of wave energy $E_w(t)$ of the main pulse.}
\label{Fig:4}
\end{figure}
\par
\indent
In summary, the propagation of the low-frequency magnetosonic pulses in a two-ion-species plasma was studied theoretically and by simulation 
with a one-dimensional
electromagnetic code based on the three-fluid model. First, the heavy-ion motion in the nonlinear pulse of the low-frequency mode was theoretically discussed. 
It was found that heavy ions gain some energy from the transverse electric field  formed in the pulse. Next, using the simulation, we showed that  the solitary pulse of the low-frequency mode is gradually damped.

\end{document}